 \documentclass[letters,fleqn,usenatbib,useAMS]{mnras}

\usepackage{psfrag,color,epsfig}

\usepackage{graphicx,graphics}	
\usepackage{amsmath}	
\usepackage{amssymb}	
\usepackage{multicol}	
\usepackage{bm}		
\usepackage{pdflscape}	
\usepackage{multirow}
\usepackage{tabularx}  

\usepackage{physics}
\usepackage{xspace}
\usepackage{soul, xcolor} 
\setstcolor{red}
\usepackage[T1]{fontenc}
\usepackage{ae,aecompl}

\usepackage{newtxmath}
\usepackage{ulem}

\def\k{\textbf{\textit{k}}}

\def\HI{\ion{H}{I}~}
\def\Hi{\ion{H}{I}}
\def\HII{\ion{H}{II}~}

\def\bxHi{\bar{x}_{\rm \ion{H}{I}}}

\def\ph{\hat{\mathbfit{p}}}

\long\def\/*#1*/{}
 
\title[Redshift Space EoR 21-cm Bispectrum]{The monopole and quadrupole moments of the Epoch of Reionization (EoR)  21-cm bispectrum}
\author[S. S. Gill et al.]{Sukhdeep Singh Gill$^{1}$\thanks{E-mail: \href{sukhdeepsingh5ab@gmail.com}{sukhdeepsingh5ab@gmail.com}},
Suman Pramanick$^{1}$,
Somnath Bharadwaj$^{1}$,
Abinash Kumar Shaw$^{2}$ and \newauthor
Suman Majumdar$^{3,4}$
\\
$^{1}$Department of Physics \& Centre for Theoretical Studies, Indian Institute of Technology Kharagpur, Kharagpur 721302, India.\\
$^{2}$Astrophysics Research Center of the Open University (ARCO), The Open University of Israel, 1 University Road, Ra’anana 4353701, Israel.\\
$^{3}$Department of Astronomy, Astrophysics and Space Engineering, Indian Institute of Technology, Indore, India\\
$^{4}$Department of Physics, Blackett Laboratory, Imperial College, London SW7 2AZ, U.K.
}

\date{\today}
\pubyear{2023}

\begin{document}
\label{firstpage}
\pagerange{\pageref{firstpage}--\pageref{lastpage}}
\maketitle


\begin{abstract}
We study the monopole ($\bar{B}^0_0$) and quadrupole ($\bar{B}^0_2$) moments of the  21-cm bispectrum (BS) from EoR simulations and present results for squeezed and stretched triangles. Both $\bar{B}^0_0$ and $\bar{B}^0_2$ are positive at the early stage of EoR where the mean neutral hydrogen (\Hi) density fraction $\bxHi \approx 0.99$. The subsequent evolution of $\bar{B}^0_0$ and $\bar{B}^0_2$ at large and intermediate scales $(k=0.29$ and $0.56 \, {\rm Mpc}^{-1}$ respectively) is punctuated by two sign changes which mark transitions in the \HI distribution. The first sign flip where $\bar{B}^0_0$ becomes negative occurs in the intermediate stages of EoR $(\bxHi > 0.5)$, at large scale first followed by the intermediate scale. This marks the emergence of distinct ionized bubbles in the neutral background. 
 $\bar{B}^0_2$ is relatively less affected by this transition, and it mostly remains positive even when $\bar{B}^0_0$ becomes negative. 
The second sign flip, which affects both $\bar{B}^0_0$ and $\bar{B}^0_2$, occurs at the late stage of EoR $(\bxHi < 0.5)$. This marks a transition in the topology of the \HI distribution, after which we have distinct \HI islands in an ionized background. This causes $\bar{B}^0_0$ to become positive. The negative  $\bar{B}^0_2$ is a  definite indication that the \HI islands survive only in under-dense regions. 
\end{abstract}



\section{Introduction}
\label{sec:intro}
The Epoch of Reionization (EoR), when the first luminous objects ionized the neutral hydrogen (\HI) gas in the intergalactic medium (IGM), is one of the least understood era in the evolution of our Universe. Observations of the  Cosmic Microwave Background (CMB) (e.g. \citealt{Planck_2016}), the absorption spectra of quasars (e.g. \citealt{Becker_2001}), and the luminosity function of Lyman-$\alpha$ emitters (e.g. \citealt{Trenti_2010})  all indirectly probe the EoR, however, these provide rather limited information about the EoR. Observations of the redshifted \HI 21-cm radiation are a very promising probe of the EoR (e.g. \citealt{Furlanetto_2006}).  Despite continued efforts using several telescopes, a
detection of the EoR 21-cm power spectrum (PS) is still forthcoming.  The main impediment for detecting the 21-cm PS  comes from foregrounds, which are $\sim 4-5$ order of magnitude stronger \citep{Ali_2008_foreground}. 
At present, the best upper limits on the EoR 21-cm PS come from the HERA experiment \citep{hera_2022}. It is anticipated that more sensitive observations with existing instruments and also the forthcoming SKA1-Low \citep{Koopmans_2015} will result in a detection.

Much of the effort to quantify the EoR 21-cm signal has focused on the 21-cm PS.  While this completely describes Gaussian fields, an early paper \citep{BP_2005} predicts the EoR 21-cm signal to be highly non-Gaussian and also provides predictions for the 21-cm bispectrum (BS), which is the lowest order statistics that captures this non-Gaussianity.  The 21-cm BS holds the potential to offer valuable insights into the evolution of the \HI distribution during the EoR \citep{Yoshiura_2015, Majumdar_2018,Majumdar_2020, Hutter_2019}, tighten the constraints on EoR models \citep{watkinson_2022,tiwari_2022} and probe the IGM physics \citep{Watkinson_2021,Kamran_2021b, Kamran_2021,  kamran_2022}.

\begin{figure*}
    \centering
    \includegraphics[width=\textwidth]{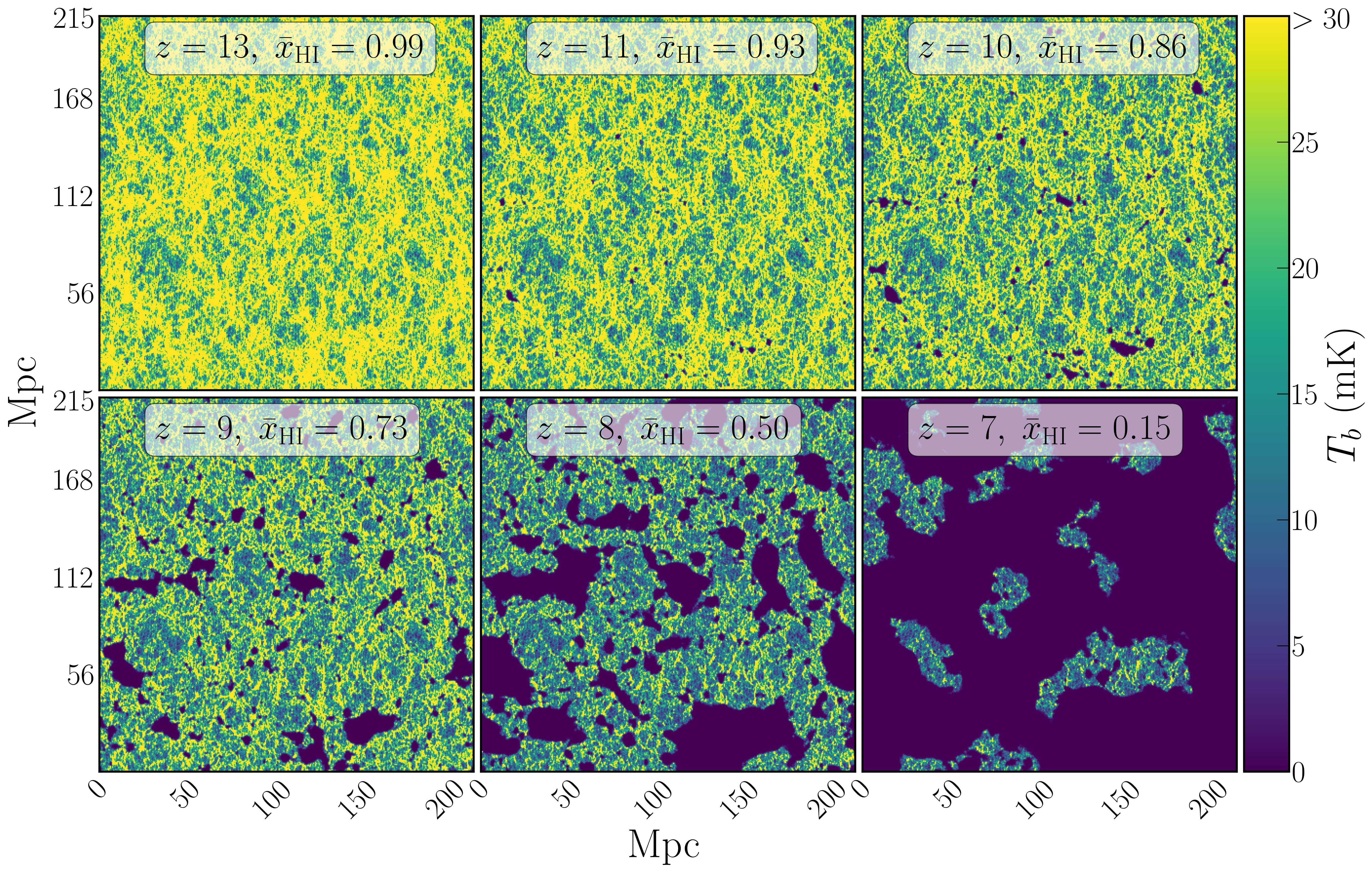}
   \caption{Evolution of the redshift space EoR 21-cm brightness temperature maps. This shows slices from a single realization of the simulated signal during different stages of EoR.
   }        
   \label{fig:Tb}
\end{figure*}

Redshift space distortion (RSD) is an important effect imprinted in the 21-cm signal \citep{BNS2001,Bharadwaj_2004b}, causing an anisotropy in the 21-cm PS along the line of sight (LoS) direction. \citet{Majumdar_2013} have quantified this anisotropy using the quadrupole moment of the 21-cm PS estimated from semi-numerical simulations of the EoR. Considering the quadrupole to monopole ratio, they show that at large scales this becomes negative when the mean \HI density fraction $\bar{x}_{\HI}$ drops to $\bar{x}_{\HI} \le  0.7$. They propose this as a direct signature of the inside-out reionization scenario. Analytical predictions (eg. \citealt{Scoccimarro_1999,Bharadwaj_2020})  suggest that the anisotropy in the BS can similarly be quantified by decomposing it into angular multipoles using spherical harmonics. In a recent study with the simulated signal, \citet{Majumdar_2020} have shown that RSD can change the amplitude of the monopole  21-cm BS  by $50-200 \%$, and can also cause it to flip sign. To the best of our knowledge, previous studies of the 21-cm BS have all been restricted to the monopole moment.  In this paper, we present predictions for the quadrupole moment of the EoR 21-cm BS using a large set of statistically independent simulated signal realizations. As the quadrupole moment is entirely due to the line of sight effects, it can potentially measure and quantify the impact of these effects on the non-Gaussian nature of the \HI distribution. Here, we study the quadrupole moment due to RSD. Moreover, the quadrupole moment of the BS used in conjunction with the monopole can impose tighter constraints on the EoR model (and its parameters) in comparison to those achievable using the monopole alone.

A brief outline of the paper follows. Section~\ref{sec:sim} presents the Methodology, Section~\ref{sec:res} presents the Results, and Section~\ref{sec:dis} presents the Discussion and Conclusions.
This paper has used the Plank+WP best-fit values of the cosmological parameters \citep{collaboration2020planck}.

\section{Methodology}
\label{sec:sim}
We have simulated the EoR 21-cm signal using the semi-numerical code ReionYuga\footnote{\href{https://github.com/rajeshmondal18/ReionYuga}{https://github.com/rajeshmondal18/ReionYuga}} \citep{ReionYuga_2021}, whose details have also been presented in \citet{Majumdar_2013} and which closely follows the homogeneous reionization scheme of  \citet{Choudhury_2009a}. The dark matter only PM N-body simulation  considers a $(215.04~{\rm Mpc})^3$  comoving volume with a $3072^3$ grids and $1536^3$ particles each of mass $1.09\times 10^8\: {\rm M}_{\sun}$. Dark matter halos were identified as having a minimum mass of $1.09\times 10^9 {\rm M}_{\sun}$ (10 particles) 
using FoF algorithm with linking length
$0.2$ grid units. The \HI reionization field was calculated on a coarser $384^3$ grids of spacing $0.56 \, {\rm Mpc}$ (comoving). The reionization model has three free parameters namely $(M_{\rm min}$, $N_{\rm ion}$, $R_{\rm mfp} )$ for which we have used the values $(1.09\times 10^9~{\rm M}_{\sun}, 23.21, 20 \, {\rm Mpc})$ respectively. Here 
$M_{\rm min}$ is the minimum halo mass assumed to host a reionization source (UV luminous galaxy), $N_{\rm ion}$ is a proportionality constant that linearly relates the number of ionizing photons emitted by an ionizing source to the amount of hydrogen contained in the host halo and $R_{\rm mfp}$ is the mean free path of the ionizing photons. This choice of simulation parameters corresponds to a model of EoR where reionization starts around $z=13$ and ends around $z=6$, marking $50\%$ ionization at $z\approx 8$. The integrated Thomson scattering optical depth computed for our fiducial model is $\tau = 0.057$, which is consistent with  $\tau = 0.054 \pm 0.007$ obtained from CMB measurements \citep{collaboration2020planck}. These fiducial simulations have been used in many previous studies \citep{Mondal_2015a,  mondal_2017, Mondal_2018,Majumdar_2018, Majumdar_2020, Shaw_2019, pramanick2023quantifying}. Varying the model parameters will change the evolution history of the mean neutral hydrogen fraction $\bxHi$. However, we do not expect the overall topology of the 21-cm signal at different stages of reionization to change drastically. The reader is referred to \citet{Shaw_2020} for a detailed discussion on the effect of varying these parameters. We have generated snapshots of the redshift space 21-cm signal at six different redshifts $z =  13, 11, 10, 9, 8$ and $7$. We have run $45$ statistically independent realizations of the simulations which provide the mean values and the error bars for the results presented here.

\begin{figure}
\centering 
\includegraphics[width=.49\textwidth]{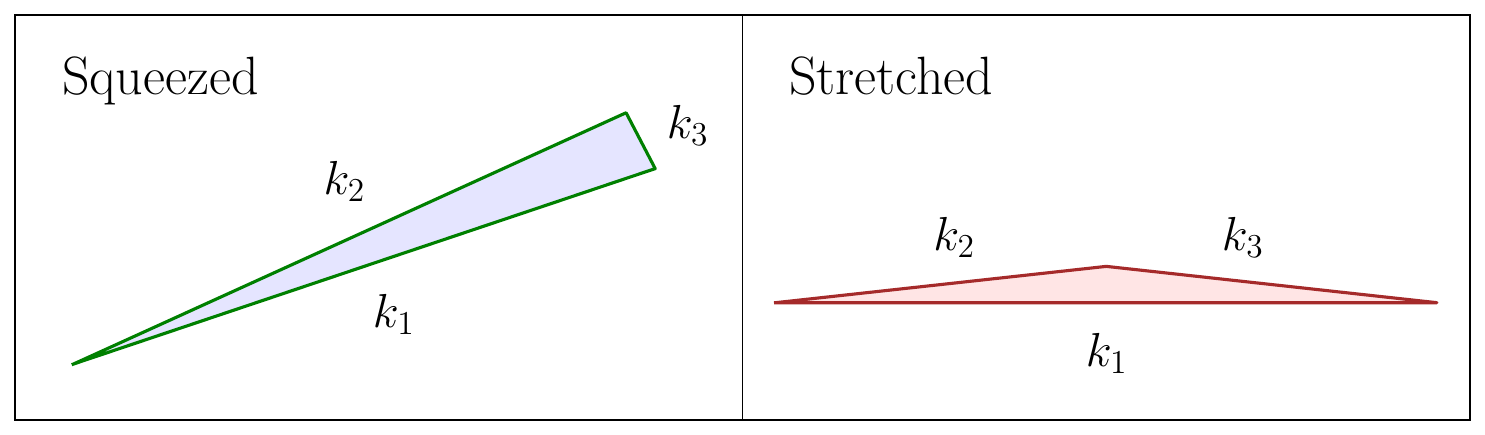} 
\caption{ This shows a squeezed triangle and a stretched triangle, the two triangle shapes for which the bispectrum results are presented in Figure~\ref{fig:bmom}. }
\label{fig:tri}
\end{figure}

The EoR 21-cm redshift space BS $B^s(\k_1,\k_2,\k_3)$ has  been   decomposed into  angular multipoles $\bar{B}_\ell^m(k,\mu,t)$ using spherical harmonics $Y_{\ell}^m(\ph)$  where 
\begin{equation}
\bar{B}^{m}_{\ell}(k,\mu,t) = \sqrt{\frac{(2 \ell +1)}{4 \pi}}
\int [Y^m_{\ell}(\ph)]^{*}  B^s(k,\mu,t,\ph) \, d \Omega_{\ph} \,.
\label{eq:bispec_int}
\end{equation}
With reference to a triangle $(\k_1,\k_2,\k_3)$,  $k$  (the largest side) quantifies its size, the two dimensionless parameters $(\mu,t)$ with $0.5 \le \mu,t \le 1$ and $2 \mu t \ge 1$  quantify its shape and the unit vector $\ph$  quantifies its orientation with respect to the LoS \citep{Bharadwaj_2020}. 
Our entire analysis is restricted to the monopole $(\ell=0)$ and quadrupole $(\ell=2)$ with $m=0$. \citet{shaw2021fast} have recently proposed a fast estimator to quantify the shape dependence of the bispectrum monopole. Here, we have slightly modified this and used it to estimate the binned bispectrum monopole and quadrupole moments for triangles of all possible unique shapes. The monopole \citep{Majumdar_2020,Mondal_2021,tiwari_2022} and the quadrupole are both found to have the largest values and the highest statistical significance (signal-to-noise ratio) for linear triangles, which corresponds to the situation where the two largest sides of the triangle are nearly aligned  in the same direction and the cosine of the angle between these two sides $\mu \approx 1$. Results are presented here only for linear triangles of  two specific shapes $(\mu,t)\approx(1.0, 1.0)$ and $(1.0, 0.5)$ which respectively correspond to squeezed  triangles $(k_1 \approx k_2, k_3 \rightarrow 0)$ and stretched  triangles $(k_1/2 \approx k_2 \approx k_3)$ both of which are  illustrated in Figure~\ref{fig:tri}. 

\begin{figure*}
		\centering 
		\includegraphics[width=\textwidth]{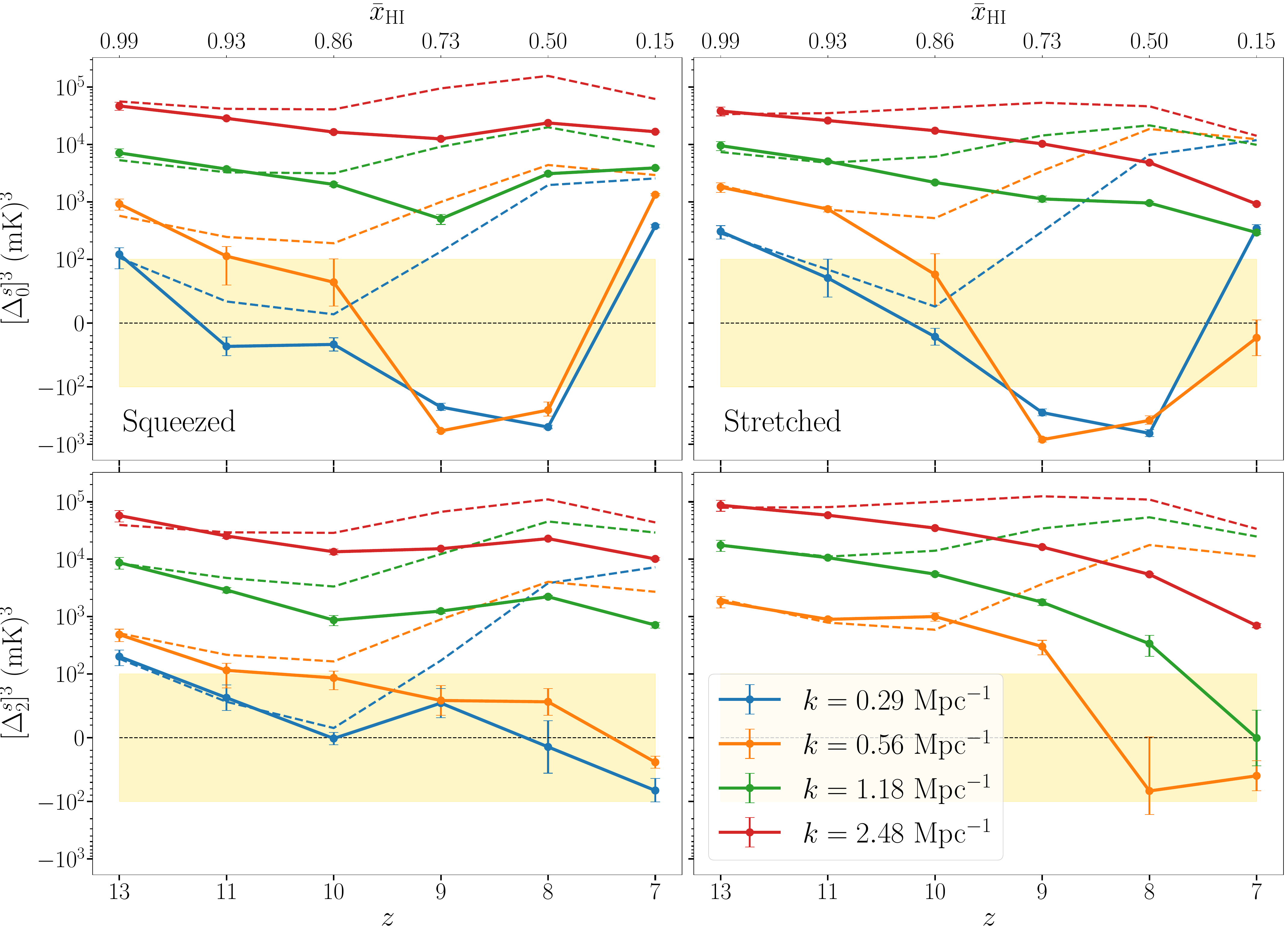} 
		\caption{The solid lines show the redshift and scale dependence of the simulated bispectrum multipole moments. We show monopole $[\Delta^s_0]^{3}$ (top row) and quadrupole $[\Delta^s_2]^{3}$ (bottom row) in terms of mean cubed brightness temperature fluctuations.
  The left and right columns correspond to squeezed and stretched triangles, respectively. The analytical predictions from second-order standard perturbation theory (2PT) are shown by dashed lines. The yellow-shaded region indicates the range where the $y$ scale is linear.}
		\label{fig:bmom}
\end{figure*}
\section{Results}\label{sec:res}
Figure \ref{fig:Tb} visually depicts the redshift evolution of the 21-cm brightness temperature distribution. At the early stage  ($z=13$), the \HI is nearly completely neutral with $\bxHi=0.99$. There are a few small isolated ionized ($\HII$) bubbles that cannot be visually identified. These isolated $\HII$ bubbles grow and merge during the intermediate stages of EoR ($z=11, 10, 9$)  where $\bxHi < 0.5$. The mid-stage ($\bxHi\approx 0.5$), which occurs at $z=8$, marks a transition in the \HI distribution. Subsequently, at the late stage $(z=7)$, the bulk of the \HI is ionized ($\bxHi=0.15$), and we are left with isolated \HI islands. The emergence of large \HII  bubbles during the intermediate and mid stages and \HI islands at the late stage makes the EoR 21-cm signal highly non-Gaussian. 

The top  row of Figure~\ref{fig:bmom} shows the redshift evolution  of 
 $[\Delta^s_0]^{3}= k^6 \bar{B}^0_0(k,\mu,t) /(2 \pi^2)^2$ (redshift space monopole) 
which we may interpret as the mean cubed brightness temperature fluctuation, and the lower row shows $[\Delta^s_2]^{3}$ (redshift space quadrupole), which has been similarly defined using  $\bar{B}^0_2(k,\mu,t)$. The left and right columns correspond to squeezed and stretched triangles, respectively. The different curves in each panel correspond to  $k=0.29, 0.56, 1.18$ and $2.48 \, {\rm Mpc}^{-1}$, which we refer to as large, intermediate, small, and smallest scales, respectively.  We have also shown (dashed lines) the predictions from second-order perturbation theory (2PT) using the formulae from \citet{Mazumdar_2020} with $P^r(k)$ being the real space 21-cm PS  estimated from the simulations as input.  The 2PT predictions assume that the 
 \HI traces the underlying matter distribution.  Due to the limited simulation volume,  we do not obtain a statistically significant (SNR $>3$) estimate of $[\Delta^s_2]^{3}$  for stretched triangles at large scales, and we have not shown this here. Considering the overall behaviour, we see that in most cases, the results for squeezed and stretched triangles are very similar, and we discuss these together.  
 
{\em The early stage of EoR $(z=13)$.} Here, we see that both  $[\Delta^s_0]^{3}$  and $[\Delta^s_2]^{3}$ are positive, and the value of the quadrupole is comparable to that of the monopole. In some cases, the quadrupole exceeds the monopole by a factor in the range of $2$  to  $3$.  We further see that  at all scales $[\Delta^s_0]^{3}$ and $[\Delta^s_2]^{3}$ closely follow the 2PT predictions $[\Delta^s_0]_{\rm 2PT}^{3}$ and $[\Delta^s_2]_{\rm 2PT}^{3}$ respectively. This indicates that it is reasonable to assume that at this stage the \HI traces the underlying matter ($M$) distribution, whereby  $[\Delta^s_0]^3 \approx [\Delta^s_0]_M^{3}$ and  $[\Delta^s_2]^3 \approx [\Delta^s_2]_M^{3}$ which arise nearly entirely from the non-Gaussianity in the underlying matter distribution. Note that we have also used 
$[\Delta^s_0]_{\rm 2PT}^{3}$ and $[\Delta^s_2]_{\rm 2PT}^{3}$  as proxies for $ [\Delta^s_0]_M^{3}$ and $ [\Delta^s_2]_M^{3}$ respectively at lower redshift. While this is not strictly correct once there is significant reionization, we still expect this to correctly capture some of the salient features. In particular, we see that both $[\Delta^s_0]_M^{3}$ and $[\Delta^s_2]_M^{3}$ increase by several orders of magnitude as we go from large scale to the smallest scale (increasing $k$), a behaviour we may also expect to hold at lower $z$. In addition, we also expect both $[\Delta^s_0]_M^{3}$ and 
$[\Delta^s_2]_M^{3}$ to scale as $\bxHi^3$ as reionization proceeds.

{\em The intermediate and mid stages of EoR $(z=11-8)$.}
We find that  both $[\Delta^s_0]^{3}$ and $[\Delta^s_2]^{3}$ deviate from  the 2PT predictions during  these stages. We attribute these deviations to the emergence of ``extra ionization ($I$) features"  in the \HI distribution and use $[\Delta^s_0]^3 \approx [\Delta^s_0]_M^{3} + [\Delta^s_0]_I^{3}$ 
to qualitatively interpret the results for the {monopole}. We note that this simple model ignores several `matter-ionization' cross-terms. However, earlier studies \citep{Majumdar_2018, Majumdar_2020} have shown that these cross-terms make a sub-dominant contribution {to the monopole}, and the two dominant terms considered here provide a reasonable qualitative understanding. As reionization proceeds, in all cases, $[\Delta^s_0]^{3}$ falls below $[\Delta^s_0]_{\rm 2PT}^{3}$, and it continues to decline to become negative at large and intermediate scales. The sign-flip to negative values occurs at the large scale first, followed by the intermediate scale, and {for both these length-scales and for both squeezed and stretched triangles}, $[\Delta^s_0]^{3}$ has particularly large negative values in the redshift range $z=8-9$. For the small and smallest scales, $[\Delta^s_0]^{3}$ remains positive throughout. 

The 21-cm signal during the intermediate and mid stages of EoR may be thought to originate from a distribution of distinct ionized bubbles embedded in a neutral medium (Figure \ref{fig:Tb}). \citet{BP_2005} have calculated the bispectrum for the distribution of non-overlapping ionized spheres in a neutral medium, and this is predicted to be negative.   Further, $\mid  [\Delta^s_0]_I^{3} \mid$  is also predicted to increase as $k^6$  for small $k$, and remain constant or decline rapidly for $k \ge \pi/R$, where $R$ is the typical bubble radius (which varies with $z$). During these stages where $\bxHi \ge 0.5$,  we also  expect  $\mid  [\Delta^s_0]_I^{3} \mid$  to increase as reionization proceeds. 
These considerations lead to a picture where the negative values of $[\Delta^s_0]^3$ arise due to the emergence of the ionized bubbles whereby $\mid  [\Delta^s_0]_I^{3} \mid  >  [\Delta^s_0]_M^{3} $ at large and mediate scales. As noted earlier, the value of $ [\Delta^s_0]_M^{3}$ increases  with $k$ and it exceeds that of 
 $\mid  [\Delta^s_0]_I^{3} \mid$ at the small and smallest scales where $[\Delta^s_0]^3$ is positive throughout.  Several earlier works \citep{Majumdar_2018,Majumdar_2020,Hutter_2019}  
 have proposed that the sign-flip from positive to negative values seen at large and intermediate scales is a distinct signature of the onset of reionization.

We now consider $[\Delta^s_2]^{3}$ during {the intermediate and mid stages of EoR}. {Similar to $[\Delta^s_0]^{3}$,  the values of $[\Delta^s_2]^{3}$ also deviate from the 2PT predictions, and these deviations mostly increase as reionization proceeds. However, unlike $[\Delta^s_0]^{3}$, we cannot use $[\Delta^s_2]^3 \approx [\Delta^s_2]_M^{3} + [\Delta^s_2]_I^{3}$ to qualitatively interpret the evolution.   It is important to note that $[\Delta^s_2]^3$  arises  due to RSD, and $[\Delta^s_2]^3$  is predicted to be zero in the absence of RSD. The peculiar velocities responsible for  RSD are sourced by fluctuations in the underlying matter distribution, and a reference to the matter fluctuations is unavoidable for modeling $[\Delta^s_2]^3$. 
Here we use $[\Delta^s_2]^3 \approx [\Delta^s_2]_M^{3} + [\Delta^s_2]_{IM}^{3}$ 
to qualitatively interpret the results for the quadrupole. The term 
$[\Delta^s_2]_{IM}^{3}$ encapsulates many possible matter-ionization cross-correlations which arise due to the effect of peculiar velocities on the ionization field. Notable among these is a term which arises from the rearrangement of the ionization field due to the peculiar velocities, a similar term for the matter field also arises in the 2PT predictions \citep{Scoccimarro_1999}.}  

 {We first consider the large scale for squeezed triangles. Here  we see that  the value of $[\Delta^s_2]^3$ matches the 2PT prediction $[\Delta^s_2]_{\rm 2PT}^{3}$   at  $z=11$. We also notice that  $[\Delta^s_2]^{3}\approx 2~|[\Delta^s_0]^{3}|$ where  $[\Delta^s_0]^{3}$ is negative at $z=11$. This indicates that at  $z=11$ we already have ionized bubbles which cause $[\Delta^s_0]_I^{3}$ to become negative and exceed $[\Delta^s_0]_M^{3} $ in magnitude. However, this does not appear to have much impact on $[\Delta^s_2]^3$ implying that $[\Delta^s_2]_{IM}^{3} \ll [\Delta^s_2]_M^{3}$. At lower $z$ $(10 \le z \le 8)$, the values of $[\Delta^s_2]^3$ drop below those of $[\Delta^s_2]_{\rm 2PT}^{3}$  and  they are also small in comparison to the monopole $([\Delta^s_2]^{3} \le 0.3 \mid [\Delta^s_0]^{3} \mid)$ which is negative.}

 {
 We next consider the intermediate scales. Here 
  $[\Delta^s_0]^{3}$ remains  positive at $z=11$ and $10$ for both squeezed and stretched triangles. For squeezed triangles $[\Delta^s_2]^{3}$  is  below  $[\Delta^s_2]_{\rm 2PT}^{3}$, however the values are quite large with  $[\Delta^s_2]^{3} \approx [\Delta^s_0]^{3}$ and $\approx 2 \, [\Delta^s_0]^{3}$ at $z=11$ and $10$ respectively. For stretched triangles,  
 $[\Delta^s_2]^{3}$  matches   $[\Delta^s_2]_{\rm 2PT}^{3}$
 and   $\approx [\Delta^s_0]^{3}$ at $z=11$, whereas $[\Delta^s_2]^{3}$  slightly exceeds    $[\Delta^s_2]_{\rm 2PT}^{3}$ and  $\approx 10 \,  [\Delta^s_0]^{3}$ at $z=10$. These results  seem to indicate that we have $[\Delta^s_0]_I^{3}< [\Delta^s_0]_M^{3}$  and $[\Delta^s_2]_{IM}^{3}< [\Delta^s_2]_M^{3}$ at $z=11$ and $10$, although the reason why   $[\Delta^s_2]^{3}$ slightly exceeds $ [\Delta^s_2]_{\rm 2PT}^{3}$ for stretched triangles at $z=10$ is not clear at present.  We see that  $[\Delta^s_0]^{3}$ drops sharply and becomes negative at lower redshifts $(z=9,8)$  for both squeezed and stretched triangles. In contrast, $[\Delta^s_2]^{3}$  declines gradually 
 and mostly remain positive in this $z$ range.  Here, the values of $[\Delta^s_2]^{3}$ are relatively small in comparison to the monopole, and they are in the range $[\Delta^s_2]^{3} \le 0.4 \mid [\Delta^s_0]^{3} \mid$.  
}

{Considering the large and intermediate scales together, the monopole  $[\Delta^s_0]^{3}$ exhibits a transition from positive to negative as reionization proceeds. In contrast ,  the quadrupole  $[\Delta^s_2]^{3}$ mostly remains positive.  Although $[\Delta^s_2]^{3}$ does become negative at $z=8$  for two cases, the error bars are large, and these values are consistent with zero. }

{
For the small and smallest scales, $[\Delta^s_2]^{3}$ roughly matches the 2PT predictions at $z=11$ and deviates from these as EoR progresses.  At these scales the values of  $[\Delta^s_2]^{3}$ remain positive throughout. For squeezed triangles, the ratio  $\mid [\Delta^s_2]^{3} / [\Delta^s_0]^{3} \mid$ has values in the range $0.5-1.5$, with the exception of $z=9$ where it shoots up to $\approx 2.5$ for small length scales.  For stretched  triangles, the ratio  $\mid [\Delta^s_2]^{3} / [\Delta^s_0]^{3} \mid$ has values in the range $1.2-2.5$, with the exception of $z=8$ where it  drops  to $\approx 0.8$ for small length scales. Considering all the length scales together, we note that the values of $\mid [\Delta^s_2]^{3} \mid$  are larger for stretched triangles in comparison to the squeezed triangles.
} 

{We reiterate that}
the quadrupole moment arises entirely due to RSD, and this is predicted to be zero in the absence of RSD. The 2PT predictions illustrate that RSD  has a strong effect on the underlying matter distribution, {as} the induced  $[\Delta^s_2]^3_M$ is positive and is comparable in magnitude to $[\Delta^s_0]^3_M$.  In contrast, RSD is expected to have a less pronounced effect on the ionized bubbles. It is possible to estimate {this} by considering the divergence of the peculiar velocities (in suitable units, eq.~(10) of \citealt{Bharadwaj_2004b}), which is comparable to the density contrast in the underlying matter distribution.  
This will be small in comparison to the {\HI  fluctuations}  produced by the ionized bubbles, and we expect $\mid [\Delta^s_2]_{IM}^{3} \mid \ll \mid  [\Delta^s_0]_{I}^{3} \mid $. This provides a qualitative interpretation for the fact that $[\Delta^s_2]^{3}$ {matches the 2PT predictions at $z=11$ whereas  $[\Delta^s_0]^{3}$ shows considerable deviations from $[\Delta^s_0]_{\rm 2PT}^{3}$  at the same redshift. This further   explains why } $[\Delta^s_2]^{3}$
remains positive at large and intermediate scales,  in contrast to $[\Delta^s_0]^{3}$, which becomes negative. {This also leads us }to  expect that $[\Delta^s_2]^{3} \approx [\Delta^s_2]^{3}_M$ at the small and smallest scales. 
 
{\em The late stage of EoR $(z=7)$.}
 Considering  the large and intermediate scales, we see that both $[\Delta^s_0]^{3}$ and 
$[\Delta^s_2]^{3}$ flip signs to respectively become positive and negative at this stage. Here  
the magnitude of  $[\Delta^s_2]^{3}$ is around $0.3 \, [\Delta^s_0]^{3}$.  At the small and smallest scales,  both $[\Delta^s_0]^{3}$  and 
$[\Delta^s_2]^{3}$ remain positive at this stage. Here the magnitude of $[\Delta^s_2]^{3}$  is comparable to that of $[\Delta^s_0]^{3}$  at the smallest scale, whereas it is approximately an order of magnitude smaller than $[\Delta^s_0]^{3}$  at the small scale. 

During the late stage of EoR, the 21-cm signal originates from individual \HI islands embedded in an ionized background (Figure \ref{fig:Tb}). This is practically a negative image of the \HI distribution during the intermediate stages of EoR, where we have individual ionized bubbles in a neutral medium, and we expect $[\Delta^s_0]^3_I$ to be similar but with an opposite sign. Considering the inside-out reionization scenario implemented here,  we expect the over-dense region to be ionized first with the \HI islands surviving mainly in the under-dense regions of the underlying matter distribution. Considering a hypothetical situation where the \HI islands reside in the over-dense regions,  RSD  would then produce a positive  $[\Delta^s_2]_{IM}^{3}$ somewhat similar to  $[\Delta^s_2]_M^{3}$. However, the peculiar velocities responsible for the RSD are reversed in the under-dense regions where the \HI islands actually reside,  and this leads to a  negative quadrupole $[\Delta^s_2]_{IM}^{3}$  during the late stage of EoR. For the large and intermediate scales,  {both $[\Delta^s_0]_M^{3}$ and $[\Delta^s_2]_M^{3}$ are expected to be quite small, and } the considerations outlined above {for $[\Delta^s_0]_{I}^{3}$ and $[\Delta^s_2]_{IM}^{3}$}
provide a qualitative interpretation for the positive and negative values of   $[\Delta^s_0]^{3}$ and  $[\Delta^s_2]^{3}$ respectively.  The magnitude of $[\Delta^s_0]_M^{3}$ and $[\Delta^s_2]_M^{3}$ both increase with $k$, and we expect these to dominate at the small and smallest scales where $[\Delta^s_0]^{3}$ and $[\Delta^s_2]^{3}$ remain positive throughout EoR. 

\section{Discussion and Conclusions}\label{sec:dis}
We have used $45$ independent realizations of the simulated  EoR 21-cm signal to obtain statistically reliable estimates for both the monopole $(\bar{B}^0_0)$ and quadrupole $(\bar{B}^0_2)$ moments of the BS, and also the $1 \sigma$ error-bars for these. Results are presented only for squeezed and stretched triangles where the BS is expected to have the largest magnitude.  To the best of our knowledge, this is the first study of the quadrupole moment of the EoR 21-cm BS. The quadrupole moment arises entirely due to RSD, and this is predicted to be zero in the absence of RSD. 

We first consider the large and intermediate scales ($k=0.29, 0.56 \, {\rm Mpc}^{-1}$ respectively) which are particularly sensitive to reionization. At the early stage, both  $\bar{B}^0_0$ and $\bar{B}^0_2$  are positive {and they match the 2PT predictions}  indicating that they trace the non-Gaussianity of the underlying matter distribution. The subsequent evolution of  $\bar{B}^0_0$ and $\bar{B}^0_2$ is punctuated by two changes in sign,  each of which marks a transition in the \HI distribution. The first sign-flip where $\bar{B}^0_0$  becomes negative occurs during the intermediate stages of EoR $(\bxHi > 0.5)$.  This sign-flip occurs first at large scales and subsequently at intermediate scales, and it marks the emergence of distinct ionized bubbles embedded in the neutral background.  It has been proposed that the negative value of $\bar{B}^0_0$ is a distinct signature of the onset of reionization \citep{Majumdar_2018,Majumdar_2020,Hutter_2019}. The ionized bubbles grow and merge as reionization proceeds.  This is reflected in the negative values of $\bar{B}^0_0$, which also grows as reionization proceeds, and we find the largest negative values near the mid-stage of EoR  $(\bxHi \approx 0.5)$. 
{In comparison to $\bar{B}^0_0$,  the quadrupole $\bar{B}^0_2$ is much less affected by the onset of reionization. In fact, in some cases, $\bar{B}^0_2$  continues to match the 2PT predictions even when $\bar{B}^0_0$ has become negative.   Further, the values of $\bar{B}^0_2$ mostly remain positive in contrast to $\bar{B}^0_0$, which becomes negative after the onset of reionization.   }

{As reionization proceeds}, the ionized regions merge,  and we are left with a distribution of distinct \HI islands embedded in an ionized background. The second sign-flip, when both $\bar{B}^0_0$ and $\bar{B}^0_2$ change signs, marks this change in the topology of the \HI distribution at the late stage of EoR  $(\bxHi < 0.5)$.  The \HI islands are somewhat like a negative image of the ionized bubbles. This causes the sign of $\bar{B}^0_0$   to change from negative to positive at the late stage. Here we note a very recent work  \citep{raste_2023}\footnote{This appeared on arxiv 
 while the present manuscript was under preparation.} which has reported a similar sign-change arising from \HI islands at the late stage of EoR. The work presented there consider two different reionization models and a variety of triangle shapes, it is however restricted to the monopole  $\bar{B}^0_0$.  Considering the quadrupole $\bar{B}^0_2$, we see that this also exhibits a sign-flip and changes from positive to negative at the late stage of EoR. The \HI islands are expected to survive only in the under-dense regions of the underlying matter distribution. This causes the peculiar velocities to be anti-correlated with the \HI distribution,  resulting in a negative $\bar{B}^0_2$.  Note that $\bar{B}^0_2$ is negative only if the \HI islands reside in under-dense regions, and it would be positive if the \HI islands were to reside in over-dense regions instead. {Hence, unlike $\bar{B}^0_0$ which only captures the transition in topology from the emergence of \HII bubbles to \HI islands, $\bar{B}^0_2$ can potentially distinguish between inside-out and outside-in models of the EoR. We note that the position-dependent power spectrum studied by \cite{Giri_2019} is also a useful tool to distinguish between such EoR models.}

The present work shows that both $\bar{B}^0_0$ and $\bar{B}^0_2$ are sensitive to changes in the topology of the \HI distribution. Considering simulations which exhibit a reionization history very similar to the present work, \citet{Bag_2019} have identified a percolation transition in the \HII distribution at $z=9$ where  $\bxHi=0.728$ and barely  $12\%$ of the volume is ionized. It is interesting to note that the BS seems to indicate a change in topology (or at least a transition) at a lower $z$,   somewhere between $z=7$ and $z=8$ where $\bxHi < 0.5$. The quadrupole moment of the large scale 21-cm PS, however, becomes negative for $\bxHi\le 0.7$ which is close to the percolation threshold.  However,  it is not clear if this behaviour of the 21-cm PS  is in anyway related to the topology of the \HI distribution. 

{The quadrupole $\bar{B}^0_2$  quantifies aspects of the non-Gaussian EoR 21-cm signal that are not captured by the monopole   $\bar{B}^0_0$.
 A joint detection  of $\bar{B}^0_0$ and $\bar{B}^0_2$ would provide a strong confirmation of the non-Gaussian behavior of  the EoR 21-cm signal. Furthermore, these two independent statistics used together would substantially enhance our understanding of reionization and 
impose tighter constraints on EoR model parameters as compared to the situation where we use  $\bar{B}^0_0$ alone.  However, the exploration of EoR model parameter estimation falls outside the scope of this work. It would also be interesting to investigate how $\bar{B}^0_0$ and $\bar{B}^0_2$ evolve under different EoR models where the topology and evolution of the 21-cm signal differ significantly.  We plan to explore this in the future.}

In conclusion, we note that  $\bar{B}^0_2$ complements $\bar{B}^0_0$ as a probe of the \HI distribution. The value of $\bar{B}^0_2$ can be comparable to that of  $\bar{B}^0_0$, and it may also exceed it by a factor in the range  $2$ to $3$ -- particularly at the small and smallest scales where both $\bar{B}^0_0$ and $\bar{B}^0_2$ are predicted to be positive throughout EoR. 
At large and intermediate scales, changes in the signs of $\bar{B}^0_0$ and $\bar{B}^0_2$ mark transitions in the \HI distribution.  In particular, we propose that a positive  $\bar{B}^0_0$  accompanied by a negative  $\bar{B}^0_2$ are a distinct and definite signature of \HI islands residing in under-dense regions at the late stage of reionization.

\section*{Acknowledgements}
SSG and SP acknowledge the support of the Prime Minister's Research Fellowship (PMRF). AKS acknowledges support by the Israel Science
Foundation (grant no. 255/18). SM acknowledges support from the Science
and Engineering Research Board (SERB), Department of Science
and Technology, GoI through the Core Research Grant No. CRG/2021/004025. We acknowledge the National Supercomputing Mission (NSM) for providing computing resources of ‘PARAM Shakti’ at IIT Kharagpur, which is implemented by C-DAC and supported by the Ministry of Electronics and Information Technology (MeitY) and Department of Science and Technology (DST), Government of India. 

\section*{Data availability}
The simulated data and the package involved in this work will be shared on reasonable request to the corresponding author.

\appendix


\bibliographystyle{mnras} 
\bibliography{ref}
\bsp
\label{lastpage}

\end{document}